\documentclass[a4paper]{article}

\usepackage{INTERSPEECH2019}
\usepackage{adjustbox}
\usepackage{multirow}
\usepackage{arydshln}
\usepackage{gensymb}
\usepackage{subfig}
\usepackage{flushend}
\usepackage{caption}
\usepackage{url}
\title{A comprehensive study of speech separation: spectrogram vs waveform separation}
\name{
Fahimeh Bahmaninezhad$^{1}$ $^\S$,
Jian Wu$^{2,4}$ $^\S$, 
Rongzhi Gu$^{3,4}$, 
Shi-Xiong Zhang$^5$, 
Yong Xu$^5$, 
Meng Yu$^5$,
Dong Yu$^5$
\thanks{$^\S$ F. Bahmaninezhad and J. Wu contributed equally to this work.}
\thanks{This work was done while F. Bahmaninezhad, J. Wu and R. Gu were interns at Tencent.}
}
\address{
 $^1$University of Texas at Dallas, Richardson TX, USA \\
 $^2$Northwestern Polytechnical University, Xi'an, China\\
 $^3$Peking University Shenzhen Graduate School, Shenzhen, China\\
 $^4$Tencent AI Lab, Shenzhen, China
 $^5$Tencent AI Lab, Bellevue WA, USA
  }
\email{fahimeh.bahmaninezhad@utdallas.edu, ,\{auszhang,lucayongxu,raymondmyu,dyu\}@tencent.com }

\begin{document}

\setlength{\abovedisplayskip}{2pt}
\setlength{\belowdisplayskip}{2pt}
\setlength{\textfloatsep}{1\baselineskip plus 0.2\baselineskip minus 0.5\baselineskip}

\maketitle
\begin{abstract}
Speech separation has been studied widely for single-channel close-talk microphone recordings over the past few years; developed solutions are mostly in frequency-domain. 
Recently, a raw audio waveform separation network (TasNet) is introduced for single-channel data, with achieving high Si-SNR (scale-invariant source-to-noise ratio) and SDR (source-to-distortion ratio) comparing against the state-of-the-art solution in frequency-domain.
In this study, we incorporate effective components of the TasNet into a frequency-domain separation method. We compare both for alternative scenarios. We introduce a solution for directly optimizing the separation criterion in frequency-domain networks.
In addition to speech separation objective and subjective measurements, we evaluate the separation performance on a speech recognition task as well.
We study the speech separation problem for far-field data (more similar to naturalistic audio streams) and develop multi-channel solutions for both frequency and time-domain separators with utilizing spectral, spatial and speaker location information.
For our experiments, we simulated multi-channel spatialized reverberate WSJ0-2mix dataset.
Our experimental results show that spectrogram separation can achieve competitive performance with better network design. Multi-channel framework as well is shown to improve the single-channel performance relatively up to +35.5\% and +46\% in terms of WER and SDR, respectively.

\end{abstract}


\vspace{-7pt}
\section{Introduction}
\label{intro}

Speech separation refers to the task of extracting all overlapping speech sources in a given mix speech signal \cite{kolbaek2017multitalker}. 
With growing speech activated technologies, such as Amazon Alexa and Apple Siri, the cocktail party problem \cite{haykin2005cocktail} has became more of an interest. Generally speaking, speech separation can be considered as a pre-processing step for other speech applications (e.g., speech and speaker recognition) when data has been collected with overlapping speech sources \cite{chen2018multi}.

Source and speech separation has been studied for many years \cite{schmidt2006single, stark2011source, kristjansson2006super, cooke2010monaural, divenyi2004speech, hu2013unsupervised}; however, after introducing deep clustering (DC) \cite{hershey2016deep} approach for speech separation and generally deep learning based methods \cite{kolbaek2017multitalker, wang2018alternative}, automatic speech separation performance is improving faster. 
%Utterance level permutation training (uPIT) \cite{kolbaek2017multitalker}, speech separation with unfolded iterative phase reconstruction \cite{wang2018end}, and etc \cite{yoshioka2018multi, erdogan2017deep} are among recent advancements in frequency-domain separation. 

Most of speech separation techniques are studied in frequency-domain \cite{kolbaek2017multitalker, hershey2016deep, xu2018single, wang2018alternative, weng2015deep} for close-talk single-channel recordings; and over the past few years new network structures \cite{wang2018end} or loss functions \cite{xu2018single, wang2018alternative} are introduced to increase the accuracy of these systems.
%; but achieved small gap improvements (should I say that?). 
In addition, authors in \cite{wang2018end} proposed an end-to-end framework with incorporating STFT operation in network layers and introduced unfolded iterative phase reconstruction to further improve the performance (inspired by the effectiveness of phase information in speech separation objective criteria).
%; however, still they are using MSE-based loss functions for optimizing the network parameters.
Speech separation with effective deep learning approaches are also started to develop for multi-channel data as well \cite{chen2018multi, yoshioka2018multi}.
%\cite{} are also proposed methods for improving phase; while improved phase can result in better separation in terms of SDR and Si-SNR, but we would like to see how much effective that can be on WER.
After all, recently \cite{luo2018tasnet} introduced a new solution, i.e., TasNet, for speech separation but in time-domain, which achieved an interesting performance comparing against the frequency-domain solutions.

In this paper, we investigate two approaches of speech separation; i.e., spectrogram (frequency-domain) and waveform (time-domain) separation in detail. We incorporate effective components of the TasNet into a spectrogram separation framework, such as: CNN-based separation network as well as loss function. We introduce an end-to-end spectrogram separation network with directly optimizing the Si-SNR criterion.
We study both waveform and spectrogram separations for multi-channel data as well, with introducing IPD (inter-microphone phase difference) features \cite{wang2018end}. 
We also evaluate the effectiveness of our multi-channel framework with speaker directional features, i.e., angle features, for extracting speech of a specific target speaker from the mix signal.
Eventually, in addition to speech separation objective and subjective criteria, we report WER as well; considering the fact that speech separation is introduced to benefit other speech related applications, including speech recognition.

\vspace{-7pt}
\section{Problem Setup}

In this section, we overview speech separation task in time and frequency-domain, introduce data used here for our experiments and describe our evaluation metrics.

\subsection{Speech Separation: Spectrogram vs Waveform}

In this section, we briefly describe spectrogram and waveform separation methods based on masking approaches.
Our focus in this paper is also on masking based separation.

The methods performing in frequency-domain, first analyze the given mix speech signals into magnitude and phase (with STFT - short-time Fourier transform), then separate the magnitude part into $S$ components (where $S$ is the number of overlapping individual sources in the mix signal) with the separation network. For reconstructing the separated speech signals in time-domain, the input mix signal phase along with the separated magnitudes are fed into ISTFT (inverse STFT).
In detail, for a mix signal $y$, the task is separating overlapping sources $x_s$ such as $y = \sum_{s=1:S}(x_s)$. If we represent $|Y|$ and $\angle Y$ as magnitude and phase of signal $y$, respectively; and $M$ as the mask functions learned with separation network, then for each individual source $x_s$ the estimated magnitude is $|\hat{X}_s| = M_s \odot |Y_s|$ (where $\odot$ is element-wise multiplication). Separated signals are then reconstructed with ISTFT using inputs $|\hat{X}_s|$ and $\angle Y$.
In \cite{wang2018end} authors proposed iterative phase reconstruction algorithm which has shown to perform better than using the $\angle Y$ for reconstruction of the signal.
In this paper, by lower-case letters, e.g., $y$, we mean time-domain samples, upper-letter $Y$ frequency-domain representation and $|Y|$ as the magnitude; $y_0$ also means time-domain samples for channel 0.

On the other hand, the TasNet as a time-domain speech separation replaces the STFT and ISTFT components with trainable encoder and decoder, respectively. Therefore, instead of mapping samples into the frequency-domain, it maps them into a new temporal resolution space. The encoder does not analyze time samples into magnitude and phase components, it operates on a unified representation which makes it different from the spectrogram separators.
In sum, the time-domain signal is mapped to a new domain with encoder and fed into the separation network. The separation network learns $S$ mask functions which are multiplied by the output of the encoder; the $S$ estimated separated sources are then converted back to the time-domain with decoder layer.

Throughout this paper, we use uPIT-BLSTM \cite{kolbaek2017multitalker} as our baseline system for spectrogram separation; and TasNet \cite{luo2018tasnet} for waveform separation.
 
\vspace{-7pt}
\subsection{Data}
\label{data}

We simulated a spatialized reverberant dataset derived from Wall Street Journal 0 (WSJ0)-2mix corpus, which is a well-studied dataset in monaural and multi-channel speech separation. There are 20,000, 5,000 and 3,000 multi-channel, reverberant, two speaker mixed speech in training, development and test subsets, respectively. 
The performance evaluation is all done on the test set, where all the speakers are unseen during training. We consider a 6-microphone circular array of 7-cm diameter, where speakers and the microphone-array are randomly located in the room. The two speakers and microphone array are on the same plane and all of them are at least 0.3m away from the wall. The image method is employed to simulate RIRs randomly from 3000 different room configurations with the size (length-width-height) ranging from 3m-3m-2.5m to 8m-10m-6m. The reverberation time T60 is sampled in a range of 0.05s to 0.5s. Samples with angle difference of 0-15\degree, 15-45\degree, 45-90\degree and 90-180\degree respectively account for 16\%, 29\%, 26\% and 29\% in the dataset.
From these 6-channels, we always use the first channel as input to the network representing the mix signal.

\vspace{-7pt}
\subsection{Metric}
Different metrics are introduced for evaluation of speech separation systems \cite{vincent2006performance}. We use Si-SNR and SDR to objectively measure the separation accuracy.
\vspace{-2pt}
\begin{equation}
\label{sdr}
\text{SDR} = 10 \log_{10} \frac{||x_{target}||^2} { ||e_{inter} + e_{noise} + e_{artif} ||^2 },
\end{equation}

\begin{equation}
\label{si-snr}
\text{Si-SNR} = 10 \log_{10} \frac{||x_{target}||^2} { ||e_{noise}||^2 },
\end{equation}
where $x_{target} = \frac{ <\hat{x} , x> x }{ ||x||^2 }$. $\hat{x}$ and $x$ respectively represent estimated and reference signals. For Si-SNR, the scale invariant is guaranteed by mean normalization of estimated and reference signals to zero mean \cite{luo2018tasnet}.

Quality of separated speech sources is also evaluated with PESQ \cite{rix2001perceptual}.
We also report WER (word error rate) to measure how the the alternative separation techniques can be effective in speech recognition tasks.

\vspace{-7pt}
\section{Improved Spectrogram Separation}

Traditional TF-masking (time-frequency masking)/spectrogram separation networks are different than TasNet in three main components:
1) \textit{Encode-decoder}: for spectrogram separation STFT/ISTFT is used to transform/reconstruct data into/from the frequency-domain, but in waveform separation, Conv-1d and ConvTranspose-1d are used respectively to encode and decode the data. Therefore, the latter one does not decompose the signal in magnitude and phase components (part of the gain achieved from waveform separation can be related to this fact).
2) \textit{Separation network}: the separation networks used in \cite{kolbaek2017multitalker} and \cite{luo2018tasnet} are also different. In \cite{luo2018tasnet}, authors mentioned the CNN-based structure outperforms the BLSTM-based network for their task (although, for spectrogram separation, alternative network structures are studied besides the BLSTM layers \cite{kolbaek2017multitalker, wang2018end}) 
%to the best of our knowledge the same CNN architecture applied in TasNet was not evaluated for spectrogram separation.
3) \textit{Training loss}: the loss function used for TF spectrogram masking is usually MSE (mean square error) \cite{kolbaek2017multitalker} which directly does not optimize the separation criterion. However, TasNet network is trained with Si-SNR loss function which directly optimizes the separation performance.

In this paper, we incorporate the CNN-based separation structure of TasNet into a spectrogram separation framework to examine the effectiveness of the network structure. 
In addition, we formulate the Si-SNR loss function for spectrogram separation as well; by replacing STFT and ISTFT with Conv-1d and ConvTranspose-1d using a specific kernel.
Details of these two modifications are explained in the following two subsections.

\vspace{-7pt}
\subsection{CNN- and BLSTM-Based Separation Network}

We use the BLSTM structure in \cite{kolbaek2017multitalker} as baseline and evaluate the performance of the system when n$\times$BLSTM layers are replaced with CNN-based separation network introduced in \cite{luo2018tasnet} (please refer to the original paper for details on the structure).

\vspace{-7pt}
\subsection{Si-SNR Loss Function}

The traditional uPIT (utterance level permutation invariant training) loss function uses MSE to train the model parameters; and in other TF-masking methods as well variations of MSE have been chosen as the training loss. To the best of our knowledge Si-SNR loss was not used with frequency domain frameworks before. \cite{wang2018end} implemented the STFT throughout the network layers but used phase sensitive MSE as their loss function.

At the end, speech separation performance will be reported in terms of Si-SNR and SDR. The TasNet network directly optimizes the Si-SNR value. Here, with some modifications in traditional implementation of STFT and ISTFT, we propose to employ Si-SNR loss while training the spectrogram separation.

Short-time Fourier transform can be formulated simply as a Conv-1d operation with a fixed specific kernel function, and ISTFT as well can be simply implemented with ConvTranspose-1d. We use the fixed kernel function with hamming window (similar to the traditional calculation of STFT). The equation of the STFT kernel function is defined in \cite{lorrymch}.
Therefore, with implementing traditional STFT and ISTFT with Conv-1d and ConvTranspose-1d, uPIT-MSE loss can be replaced easily with uPIT-SiSNR. For computing loss function, first need to call ConvTranspose-1d using mix signal phase, and estimate time-samples and then compute Si-SNR.

\begin{figure}
\centering
\begin{minipage}{.25\textwidth}
  \centering
  \includegraphics[width=0.85\linewidth]{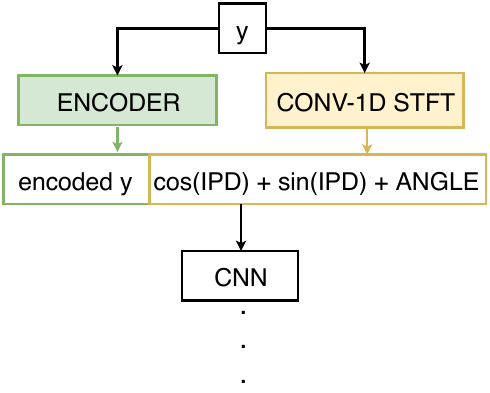}
  %\captionof{figure}{waveform separation}
  \label{fig:test1}
\end{minipage}%
\begin{minipage}{.25\textwidth}
  \centering
  \includegraphics[width=0.7\linewidth]{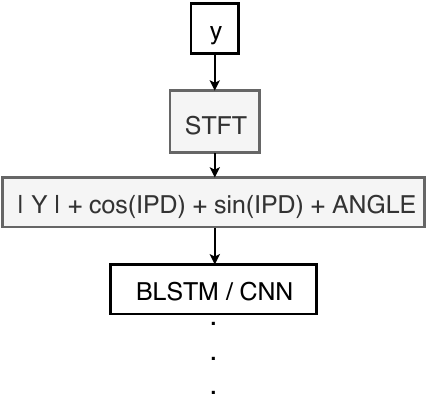}
  %\captionof{figure}{spectrogram separation}
  \label{fig:test2}
\end{minipage}
\caption{ Multi-channel speech separation; (left) waveform separation, (right) spectrogram separation.}
\label{fig:cat}
\end{figure}

\vspace{-7pt}
\section{Multi-Channel Speech Separation}
\vspace{-3pt}
As described in subsection \ref{data}, we simulated multi-channel spatialized WSJ0-2mix for our evaluations. 
In this section, the integration of multi-channel data into waveform and spectrogram separation networks is described.

In subsection \ref{sec:ipd} we introduce the spatial features, i.e., IPD, used in our multi-channel framework.
IPD (inter-microphone phase difference) features are representing the correlation of multi-channel signals and are useful for spatial locating of overlapping sources \cite{chen2018multi}.
Next, in subsection \ref{sec:angle}, angle features are explained which we used them to locate the speakers and extract the speech of a target speaker \cite{chen2018multi}.

\vspace{-7pt}
\subsection{Single-Channel vs Multi-Channel Separation}
\label{sec:ipd}

IPD (inter-microphone phase difference) features represent spatial location information \cite{chen2018multi} and are calculated as,
\begin{equation}
\label{ipd}
\text{IPD}_{i,t,f} = \angle (\frac{Y_{i_1, t,f}}{Y_{i_2, t, f}}), i=1:6
\end{equation}
where $i$ represents an entry in a microphone pair list defined for calculating the IPD; and $i_1$ and $i_2$ are the index of microphones in each pair. We calculate IPD features for 6 pairs and then concatenate 6-cos(IPD) and 6-sin(IPD) together. %For more information on IPD refer to \cite{chen2018multi}.

Figure \ref{fig:cat} shows the integration of IPD features (as well as angle features described in the next subsection) into spectrogram and waveform separation networks.
For spectrogram separation, multi-channel framework is straightforward; the IPD features are appended with $|Y_0|$, i.e., magnitude of channel 0, and then are fed into the network. 
For waveform separation, the time-samples at the input are analyzed with Conv-1d implementation of STFT and IPD features are calculated; then they are appended to the encoder output, and followed by the same separation network.
However, concatenation of IPD features in time-domain networks can be applied with various fusion methods, which is explored more in detail in \cite{lorrymch}.

\vspace{-7pt}
\subsection{Target Speaker Extraction vs uPIT Separation}
\label{sec:angle}
\vspace{-3pt}
Multi-channel separation networks can also use speakers location information to extract the target speaker of interest. Here, we use angle features \cite{chen2018multi} for this purpose.

Speech separation networks need to have a prior knowledge of number of overlapping speakers. Therefore, for different conditions when number of overlapping speakers are changing or is unknown, we need to train different models (as these systems usually estimate one mask function for each speaker simultaneously).
Using directional information or speaker-related knowledge can help to extract a target speaker; and with a single trained model and multiple passes can extract all overlapping sources.
Therefore, we extend our multi-channel framework with angle features (A) \cite{chen2018multi} to locate a target speaker and extract his/her speech; the block-diagram is shown in Figure \ref{fig:cat}. These directional features are formulated as,
\vspace{-2pt}
\begin{equation}
\label{angle}
A_{s,t,f} = \sum_{i=1}^{6} \frac{ e_s^{i,f}  \frac{ Y_{i_1, t,f}}{Y_{i_2, t,f}}  }{ | e_s^{i,f} \frac{ Y_{i_1, t,f}}{Y_{i_2, t,f}} | },
\end{equation}
where $s$ is the speaker index and $e_s^{i,f}$ represents steering vector coefficient of speaker $s$ direction of arrival at microphone $i$ for frequency $f$ \cite{chen2018multi}. 
%As mentioned earlier we calculate IPD and A for 6 pairs of microphones ...
In our experiments, we append both speaker angle features together, and always assume the first one is the target speaker for optimizing the target extraction loss.

For traditional uPIT loss function, where $S$ speakers are separating simultaneously, their reference outputs are required for optimizing the network. However, for TGT (target) loss only the reference speech of the target speaker is required.
Using more informative directional and speaker related features are studied further in \cite{lorrytgt}.

% -- I do not know why we are doing this experiment ... need more justification why this is added

\begin{table*}[!th]
\scriptsize
\caption{{ \it Experimental setup for time/frequency-domain models.}}
\centering
%\resizebox{14cm}{!}{
\begin{tabular}{c  c  c  c  c}
\toprule
\textbf{Model} & \textbf{Input} & \textbf{\# of Parameters} &  \textbf{Setting} & \textbf{Normalization}\\ \hline
F-CNN-1 & $|Y_0|$ & 8.78M & N=257 & gLN \\
F-CNN-2 & $|Y_0|$ + cos(IPD) + sin(IPD) & 9.58M & N=257$\times$13 & gLN\\
F-CNN-3 & $|Y_0|$ + cos(IPD) + sin(IPD) + Angle & 9.71M & N=257$\times$15 & gLN \\
F-BLSTM-1 & $|Y_0|$ & 67.05M & 4 $\times$ BLSTM-896 & -\\
F-BLSTM-2 & $|Y_0|$ + cos(IPD) + sin(IPD) & 89.15M & 4 $\times$ BLSTM-896 & - \\
F-BLSTM-3 & $|Y_0|$ + cos(IPD) + sin(IPD) + Angle & 92.84M & 4 $\times$ BLSTM-896 & - \\
T-CNN-1 & $y_0$ & 8.76M & L/N=40/256 & gLN \\
T-CNN-2  & $y_0$ + cos(IPD) + sin(IPD) & 8.83M & L/N=40/256 &  BN \\
\bottomrule
\end{tabular}
%}
\label{tab:setup}
\vspace{-0.25cm}
\end{table*}

% ______________________________________________________________________

\begin{table*}[!th]
\scriptsize
\caption{{ \it Comparing CNN separation network and Si-SNR loss function in spectrogram separation with single-channel input.}}
\centering
\resizebox{14cm}{!}{
\begin{tabular}{|c |c | c c c c | c | c c c c | c |}
\hline
\textbf{Model} & \textbf{Loss} & \multicolumn{5}{c|}{\textbf{Si-SNR}} & \multicolumn{5}{c|}{\textbf{SDR}}\\
&  & 0-15\degree & 15-45\degree & 45-90\degree & 90-180\degree & AVG & 0-15\degree & 15-45\degree & 45-90\degree & 90-180\degree & AVG \\
\hline
\multirow{2}{*}{F-BLSTM-1} & uPIT-SiSNR & 7.54 & 7.80 & 7.72 & 7.81 & 7.74 & 8.14 & 8.39 & 8.29 & 8.38 & 8.32 \\
 & uPIT-MSE & 6.50 & 6.79 & 6.67 & 6.78 & 6.71 & 6.98 & 7.24 & 7.11 & 7.22 & 7.16 \\
\multirow{2}{*}{F-CNN-1} & uPIT-SiSNR & 7.08 & 7.48 & 7.45 & 7.48 & 7.42 & 7.70 & 8.06 & 8.02 & 8.06 & 8 \\
 & uPIT-MSE & 6.31 & 6.67 & 6.54 & 6.65 & 6.58 & 6.81 & 7.13 & 7.0 & 7.1 & 7.04 \\
\hline
\end{tabular}
}
\label{tab:sisnr}
\vspace{-0.25cm}
\end{table*}

% ______________________________________________________________________

\begin{table*}[!th]
\scriptsize
\caption{{ \it$|Y_0|$+cos(IPD)+sin(IPD)+Angle are input to the network to evaluate the performance of multi-channel framework in extracting the speaker of interest.}}
\centering
%\resizebox{14cm}{!}{
\begin{tabular}{|c | c | c c c c | c | c c c c | c |}
\hline
\multirow{2}{*}{\textbf{Model}} & \multirow{2}{*}{\textbf{Loss}} & \multicolumn{5}{c|}{\textbf{Si-SNR}} & \multicolumn{5}{c|}{\textbf{SDR}} \\
&  & 0-15\degree & 15-45\degree & 45-90\degree & 90-180\degree & AVG & 0-15\degree & 15-45\degree & 45-90\degree & 90-180\degree & AVG \\
\hline
\multirow{2}{*}{F-BLSTM-3} & uPIT-SiSNR & 5.09 & 9.11 & 9.62 & 9.31 & 8.72 & 5.81 & 9.64 & 10.11 & 9.85 & 9.27 \\
 & TGT-SiSNR & 4.59 & 9.07 & 9.59 & 9.13 & 8.58 & 5.31 & 9.60 & 10.08 & 9.64 & 9.12 \\
\multirow{2}{*}{F-CNN-3} & uPIT-SiSNR & 5.59 & 9.38 & 10.29 & 10.79 & 9.49 & 6.92 & 10.36 & 11.2 & 11.75 & 10.50 \\
 & TGT-SiSNR & 4.88 & 9.69 & 10.32 & 9.80 & 9.2 & 5.61 & 10.19 & 10.77 & 10.27 & 9.71 \\
\hline
\end{tabular}
%}
\label{tab:tgt}
\vspace{-0.25cm}
\end{table*}

% ______________________________________________________________________

\begin{table*}[!th]
\scriptsize
\caption{{ \it Comparing spectrogram and waveform separation for both separation and ASR tasks.}}
\centering
\resizebox{17cm}{!}{
\begin{tabular}{|c c c | c c c c | c | c c c c | c | c | c |}
\hline
\multirow{2}{*}{\textbf{\# Channels}} & \multirow{2}{*}{\textbf{Domain}} & \multirow{2}{*}{\textbf{Model}} & \multicolumn{5}{c|}{\textbf{Si-SNR}} & \multicolumn{5}{c|}{\textbf{SDR}} & \multirow{2}{*}{\textbf{PESQ}} & \multirow{2}{*}{\textbf{WER Reduc. (\%)}} \\
& & & 0-15\degree & 15-45\degree & 45-90\degree & 90-180\degree & \textbf{AVG} & 0-15\degree & 15-45\degree & 45-90\degree & 90-180\degree & \textbf{AVG} &  &  \\
\hline
\multirow{4}{*}{1-ch} & \multirow{2}{*}{time} & T-BLSTM & - & - & - & - & - & - & - & - & - & - & -  & - \\
&  & T-CNN-1 & 9.02 & 9.33 & 9.59 & 9.71 & \textbf{9.47} & 9.57 & 9.83 & 10.09 & 10.2 & \textbf{9.97} & 1.95 & 45.53\\
& \multirow{2}{*}{freq} & F-BLSTM-1 & 7.54 & 7.80 & 7.72 & 7.81 & \textbf{7.74} & 8.14 & 8.39 & 8.29 & 8.38 & \textbf{8.32} & 1.77 & 32.21\\
& & F-CNN-1 & 7.08 & 7.48 & 7.45 & 7.48 & \textbf{7.42} & 7.70 & 8.06 & 8.02 & 8.06 & \textbf{8} & 1.77 & 35.17 \\
\multirow{4}{*}{m-ch} & \multirow{2}{*}{time} & T-BLSTM & - & - & - & - & - & - & - & - & - & - & -  & - \\
&  &  T-CNN-2 & 7.70 & 11.63 & 12.33 & 12.62 & \textbf{11.55} & 8.31 & 12.07 & 12.74 & 13.03 & \textbf{11.99} & 2.10 & 59.11\\
& \multirow{2}{*}{freq} & F-BLSTM-2 & 5.41 & 9.37 & 10.13 & 10.65 & \textbf{9.38} & 6.13 & 9.89 & 10.62 & 11.13 & \textbf{9.91} & 1.92 & 45.32\\
&  & F-CNN-2 & 6.88 & 10.27 & 11.02 & 11.54 & \textbf{10.36} & 7.5 & 10.75 & 11.47 & 11.99 & \textbf{10.84} & 2.00 & 51.73\\
\hline 
%\multicolumn{15}{|c|}{\textbf{Oracle Masks}}  \\
\multicolumn{3}{|c|}{Oracle IBM}    & 11.56 & 11.51 & 11.53 & 11.53 & \textbf{11.53} & 11.93 & 11.86 & 11.88 & 11.88 & \textbf{11.88} & 2.01 & 50.37\\
\multicolumn{3}{|c|}{Oracle IAM} & 11.05 & 11.03 & 11.05 & 11.03 & \textbf{11.04} & 11.33 & 11.3 & 11.31 & 11.29 & \textbf{11.30} & 2.23 & 71.45\\
\multicolumn{3}{|c|}{Oracle IRM} & 11.01 & 10.96 & 10.98 & 10.97 & \textbf{10.98} & 11.45 & 11.39 & 11.39 & 11.39 & \textbf{11.40} & 2.22 & 70.28\\
\multicolumn{3}{|c|}{Oracle IPSM} & 13.68 & 13.6 & 13.64 & 13.63 & \textbf{13.63} & 14.04 & 13.94 & 13.98 & 13.97 & \textbf{13.98} & 2.28 & 71.10\\
\hline
\multicolumn{3}{|c|}{Reference} & & & & & & & & & & & 2.35 & 73.01\\
\hline
\end{tabular}
}
\label{tab:sw}
\vspace{-0.25cm}
\end{table*}

% ______________________________________________________________________

\vspace{-7pt}
\section{Experiments}
\subsection{Experimental Setup}

Table \ref{tab:setup} summarizes the network setup and details of models we applied in our system development (F- and T- mean frequency and time-domain in the tables). 
Parameters L (length of filters in encoder), N (number of filters in encoder), X (number of convolutional blocks in each repeat), R (number of repeats), P (kernel size in convolutional blocks), B (number of channels in bottleneck 1 $\times$ 1-conv block) and H (Number of channels in convolutional blocks) have the same definition as used in \cite{luo2018tasnet}. Please refer to \cite{luo2018tasnet} for more details on the structure of the separation network and its hyper-parameters.
For all of our time-domain networks hyper-parameters X/R/B/H/P=8/4/257/512/3 are fixed, and for the rest details are included in the Table \ref{tab:setup}. The gLN and BN in Table \ref{tab:setup} are also global layer normalization and batch normalization.
In addition, the selected pairs for IPDs are (1, 4), (2, 5), (3, 6), (1, 2), (3,4) and (5, 6) in all experiments.

%We used GoogleWeb API speech recognition tool to calculate WER here. The correct permutation of separated speech sources are decided based on SDR scores; then used to calculate WER and PESQ.
We used KALDI \cite{povey2011kaldi} ASR recipe for WSJ (trained on clean non-reverberate data) to calculate the WERs. 
For reference, the mix signal WER is 89.91\%, and we report the WER reduction rate obtained with each developed speech separation system.
In addition, scores for oracle ideal binary mask (IBM), ideal amplitude mask (IAM), ideal ratio mask (IRM) and ideal phase sensitive mask (IPSM) are all reported.

\vspace{-7pt}
\subsection{Experimental Result}
\vspace{-3pt}
We evaluate our proposed solutions in three different evaluation sets.
First, incorporating CNN separation network as well as Si-SNR loss function in spectrogram separation is evaluated. Table \ref{tab:sisnr} summarizes the results. This table only includes results from single-channel models.
Results confirm the effectiveness of the Si-SNR loss function over MSE. However, for the single-channel input, BLSTM structure slightly performs better than CNN structure.

Next, for time-frequency domain masking network we employ IPD and angle features to extract a target speaker. We evaluate the effectiveness of the multi-channel framework with Si-SNR loss function, comparing BLSTM and CNN separation network structures here.
Table \ref{tab:tgt} shows the results. Here, uPIT-SiSNR and TGT-SiSNR are compared against each other. 
The results confirm the effectiveness of the proposed multi-channel framework. However, better directional features \cite{lorrytgt} may be applied here to gain better performance on target speaker extraction. 

Finally, we compare waveform separation against spectrogram separation, for both single-channel and multi-channel frameworks.
Table \ref{tab:sw} (in the table 1-ch means single channel input, m-ch means multi-channel input) shows the results including all Si-SNR, SDR, PESQ and WER scores to not only compare their performance for speech separation task, but also evaluate the impact of them on a speech recognition task. All models in this table are trained with uPIT-SiSNR loss function.
% The results confirm that using Si-SNR loss function with CNN network decreases the gap between time-domain and frequency-domain separation, specially on multi-channel input. 
The results confirm that for multi-channel (using $|Y_0|$ + cos(IPD) + sin(IPD)) models, CNN separation structure helps to decrease the gap further between time and frequency domain models.
In all tables, Si-SNR and SDR scores are reported on 4 sets of angle-differences between the two overlapping speakers as well; which results show that for smaller angle differences, more informative inputs are required.

\vspace{-7pt}
\subsection{Discussion}
\vspace{-3pt}
Based on the results summarized in Tables \ref{tab:sisnr}, \ref{tab:tgt}, \ref{tab:sw}, we can confirm the effectiveness of CNN-based separation structure in frequency-domain. Although, for single-channel inputs F-BLSTM-1 slightly outperforms F-CNN-1 in terms of SDR and Si-SNR; in terms of WER, F-CNN-1 performs better and for multi-channel framework F-CNN-2 achieves better results than F-BLSTM-2 for all criteria.
Moreover, Multi-channel models always outperform single-channel models in terms of all reported criteria; F-BLSTM-2, F-CNN-2, and T-CNN-2 relatively gain +19.11\%, +35.5\%, and +20.26\% improvement in terms of SDR criterion, respectively.
F-CNN-2 decreases the gap between frequency-domain (with using phase of the mix signal) and time-domain separators; specially in terms of PESQ, results are competitive, 2.1 for T-CNN-2 and 2 for F-CNN-2.
In sum, utilizing multi-channel framework can benefit speech separation task. Although, frequency-domain separation improved with Si-SNR and better separation network design; it can improve further with using better phase component at the reconstruction layer.

\vspace{-7pt}
\section{Conclusions}
\vspace{-3pt}
In this study, multi-channel spatialized reverberant WSJ0-2mix dataset prepared and studied for alternative speech separation setups; including single and multi-channel models, target speaker extraction framework, different loss functions for both spectrogram and waveform input representations.
We improved the performance of frequency-domain speech separation with incorporating Si-SNR loss function (directly optimizing the separation criterion). In addition, we explored replacement of BLSTM separation network with CNN structure, which improved the performance for multi-channel inputs, with almost 10 times less number of parameters.
We reported PESQ, WER, SDR and Si-SNR scores, to not only evaluate the performance of the introduced models on separation task, but also on speech recognition task as well.
Incorporating separation network and loss function from TasNet into frequency-domain shown to be effective; with using CNN-separator, SDR relatively improved by up to +9\% on multi-channel scenario; and with Si-SNR loss function SDR improved by up to +16\% for single-channel models.
With multi-channel framework, SDR, Si-SNR, PESQ and WER relatively improved over the single-channel framework by up to +35.5\%, +39.62\%, +12.99\% and +46\%, respectively.

\bibliographystyle{IEEEtran}
\bibliography{mybib}
\end{document}